\documentclass{emulateapj}
\usepackage{apjfonts}
\usepackage{graphicx}

\def\d3{$\delta_{3}$ }
\def\1d3{$(1 + \delta_{3})$ }
\def\l1d3{$\log_{10}(1 + \delta_{3})$ }
\def\dn{$\delta_{n}$ }
\def\onedn{$(1 + \delta_{n})$ }
\def\pdn{$\log_{10}(1 + \delta_{n})$ }
\def\s3{$\Sigma_{3}$}

\def\24m{24 $\mu$m}

\def\sm{$M_{*}$ }
\def\kpc{$h^{-1}$kpc }
\def\kms{${\rm km~s^{-1}}$ }

\def\dis{$r_{p}$ }
\def\vel{$|$$\Delta$$v$$|$}

\def\mbe{$M_{B}^{e}$ }
\def\abm{$M_{B} $}
\def\Nc{$N_{c}$ }
\def\Ncb{$N_{c}^{b}$ }
\def\Ncr{$N_{c}^{r}$ }
\def\Ncm{$N_{c}^{m}$ }

\def\aa{A\&A}

\shorttitle{Environment of WET, DRY, AND MIXED GALAXY MERGERS}

\shortauthors{Lin et al.}

\begin{document}

\title{Where do Wet, Dry, and Mixed Galaxy Mergers Occur? A Study of the Environments of Close Galaxy Pairs in the DEEP2 Galaxy Redshift Survey}

\author{Lihwai Lin \altaffilmark{1}, Michael C. Cooper \altaffilmark{2,3}, Hung-Yu Jian \altaffilmark{4}, David C. Koo \altaffilmark{5}, David R. Patton \altaffilmark{6}, Renbin Yan \altaffilmark{7}, Christopher N. A. Willmer \altaffilmark{2}, Alison L. Coil \altaffilmark{8}, Tzihong Chiueh \altaffilmark{4}, Darren J. Croton \altaffilmark{9}, Brian F. Gerke \altaffilmark{10}, Jennifer Lotz \altaffilmark{11,12}, Puragra Guhathakurta \altaffilmark{5}, and Jeffrey A. Newman \altaffilmark{13}}

\altaffiltext{1}{Institute of Astronomy \& Astrophysics, Academia Sinica, Taipei 106, Taiwan; Email: lihwailin@asiaa.sinica.edu.tw}
\altaffiltext{2}{Steward Observatory, University of Arizona, 933 N.\ Cherry Avenue, Tucson, AZ 85721 USA}
\altaffiltext{3}{Spitzer Fellow}
\altaffiltext{4}{Department of Physics, National Taiwan University, Taipei, Taiwan}
\altaffiltext{5}{UCO/Lick Observatory, Department of Astronomy and Astrophysics, University of California, Santa Cruz, CA 95064}
\altaffiltext{6}{Department of Physics and Astronomy, Trent University, 1600 West Bank Drive, Peterborough, ON K9J 7B8 Canada}
\altaffiltext{7}{Department of Astronomy and Astrophysics, University of Toronto, 50 St. George Street, Toronto, ON M5S 3H4, Canada}
\altaffiltext{8}{Department of Physics and Center for Astrophysics and Space Sciences, University of California, San Diego, 9500 Gilman Dr., La Jolla, CA 92093}
\altaffiltext{9}{Centre for Astrophysics \& Supercomputing, Swinburne University of Technology, P.O. Box 218, Hawthorn, VIC 3122, Australia}
\altaffiltext{10}{Kavli Institute for Particle Astrophysics and Cosmology, Stanford Linear Accelerator Center, 2575 Sand Hill Rd., M/S 29, Menlo Park, CA 94025, USA}
\altaffiltext{11}{National Optical Astronomy Observatory, 950 N. Cherry Ave., Tucson, AZ 85719}
\altaffiltext{12}{Leo Goldberg Fellow}
\altaffiltext{13}{Physics and Astronomy Dept., University of Pittsburgh, Pittsburgh, PA, 15620}

\begin{abstract}
\hspace{3mm} We study the environment of wet, dry, and mixed galaxy mergers at $0.75 < z < 1.2$ using close pairs in the DEEP2 Galaxy Redshift Survey,
aiming to establish a clear picture of how the cosmic evolution of various merger types relate to the observed large-scale extra-galactic environment and its role in the growth of red-sequence galaxies. We find that the typical environment of mixed and dry mergers is denser than that of wet mergers, mostly due to the color-density relation. While the galaxy companion rate (\Nc) is observed to increase with overdensity, using N-body simulations we find that the fraction of pairs that will eventually merge decreases with the local density, predominantly because interlopers are more common in dense environments.
After taking into account the merger probability of pairs as a function of local density, we find only marginal environment dependence of the fractional merger rate for wet mergers over the redshift range we have probed. On the other hand, the fractional dry merger rate increases rapidly with local density due to the increased population of red galaxies in dense environments. In other words, while wet mergers transform galaxies from the blue cloud into the red sequence at a similar fractional rate across different environments (assuming that the success rate of wet mergers to yield red galaxies does not depend on environment), the dry and mixed mergers are most effective in overdense regions. We also find that the environment distribution of K+A galaxies is similar to that of wet mergers alone and of wet+mixed mergers, suggesting a possible connection between K+A galaxies and wet and/or wet+mixed mergers.
Based on our results, we therefore expect that the properties, including structures and masses, of red-sequence galaxies should be different between those in underdense regions and in overdense regions since the dry mergers are significantly more important in dense environments.
We conclude that, as early as $z \sim 1$, high-density regions are the preferred environment in which dry mergers occur, and that present-day red-sequence galaxies in overdense environments have, on average, undergone 1.2$\pm$0.3 dry mergers since this time, accounting for (38$\pm$10)\% of their mass accretion in the last 8 billion years. Our findings suggest that dry mergers are crucial in the mass-assembly of massive red galaxies in dense environments, such as Brightest Cluster Galaxies (BCGs) in galaxy groups and clusters.
\end{abstract}


\section{INTRODUCTION}

Within the framework of hierarchical structure formation, dark-matter
halos grow through successive mergers with other halos and through
accretion of the surrounding mass \citep{blu84,dav85,ste09}. Whether this bottom-up scenario also holds for galaxies which reside in dark matter halos remains a challenging question in the theories of galaxy formation and evolution. The keys to pin down the importance of mergers in the assembly history of galaxies are the study of galaxy merger rates as a function of cosmic time \citep{car00,pat02,con03,lin04,lin08,lot08a,der09,blu09} and to understand the level of triggered star formation during galaxy interactions \citep{lam03,nik04,woo06,lin07,bar07,ell08}.

In addition to assembling galaxy masses, galaxy mergers have also been suggested to be responsible for the change of galaxy properties \citep{hop06}. Galaxy populations have been shown to evolve differently since redshift 1.5: while the characteristic number densities of blue galaxies remain fairly constant, the number and stellar mass densities of red galaxies have at least doubled over this period \citep{bel04,wil06,fab07}. The growth rate of the red galaxies is much faster than the predictions from purely passive evolution, suggesting that additional physical mechanisms are required to truncate the star formation in some of the blue galaxies and turn them into the red-sequence \citep{bel07}. More recently, there have been studies examining the connection between galaxy mergers and the establishment of red-sequence galaxies \citep{van05,bel06a,lin08,ske09}. Using close pairs found in the DEEP2 Survey, \citet{lin08} found that the present red galaxies might have experienced on average 0.7, 0.2, and 0.4 wet, dry, and mixed mergers respectively since $z \sim 1$, suggesting a key role of galaxy mergers in the evolution history of red galaxies. In addition, the relative role of different types of mergers also evolve with redshift, indicating that the effect of quenching star formation and mass build-up through mergers also evolve with time. Meanwhile, it becomes increasingly clear that dense regions such as galaxy groups and clusters might be the places where the transformation of blue galaxies into red galaxies occurs most effectively. If galaxy mergers are the dominant quenching mechanism, one should also expect a clear environment dependence of galaxy merger rates. However, there have been very few attempts to probe the connection between mergers and environment observationally \citep{mci08,dar09}.

Another way to probe the connection between mergers and the formation
of red galaxies is to compare the environment distribution of mergers
to the poststarburst galaxies, the so-called 'K+A' or 'E+A' galaxies
\citep{dre83}. Such galaxies are identified through their strong
Balmer absorption and little H$\alpha$ or [OII] emissions, indicating
that they had recent star formation with the last $\sim 1$ Gyr or so,
but no on-going star formation. These K+A galaxies are suggested to be
the transition phase between star forming galaxies and the dead
red-sequence galaxies, and hence they could be the direct progenitors
of early-type red galaxies. There have been many studies looking at
the environment of poststarburst galaxies
\citep{got05,hog06,yan09,pog08}, with the aim of identifying the
process that truncates the star formation activity (mergers,
ram-pressure stripping, AGN feedback, etc.). By comparing the
environment distribution of K+A galaxies in the literature to that of
galaxy mergers, we can gain insight on the importance of galaxy
mergers in the formation of K+A galaxies and hence the built-up of red-sequence galaxies.

Since galaxy interactions require more than one galaxy by definition,
it is expected that galaxy mergers tend to reside in dense
regions. Galaxy groups are thought to be preferred places for galaxy
mergers because of their lower velocity dispersions as opposed to the
galaxy clusters. By studying four X-ray luminous groups at
intermediate redshifts ($z \sim 0.4$), \citet{tra08} suggested that
dry mergers are an important process to build up massive galaxies in
the cores of galaxy groups/clusters; \citet{mci08}, using group and
cluster samples in SDSS also found that the frequency of mergers
between luminous red galaxies (LRGs) is significantly higher in groups
and clusters compared to overall population of LRGs. While most of
previous studies examining the environment of mergers focused on the
dry mergers in dense environments, to date no quantitative measurement
of wet, dry, and mixed merger rates as a function of environment has
been obtained. This paper aims to address the issue of where galaxies build up their masses and where the transformation of galaxies happens by exploring the environment of various types of interacting galaxies. There are two methods that have been used in DEEP2 to classify environments: one is to use the projected $n^{th}$-nearest neighbor surface density $\Sigma_{n}$ \citep{coo06}, which gives the estimates of local density of individual galaxies; the other is to classify the galaxy environments into "field" and "groups/clusters" \citep{ger05}. In this work, we adopted the former approach as a primary environment measurement to investigate (1) which environment hosts most wet/dry/mixed mergers and (2) the pair fraction and the fractional merger rate as a function of environment at $0.75 < z < 1.2$, using the blue-blue, red-red, and blue-red pairs selected from the DEEP2 sample \citep{lin08}. It is worth noting that there are potential caveats in our analysis using galaxy colors to classify wet/dry/mixed mergers. At $z \sim 1$, it is found that about 20\% of red galaxies appear to be either edge-on disks or dusty galaxies and hence are likely to be gas-rich \citep{wei05} whereas there also exist blue spheroidals that could be gas-poor, although these are relatively rare objects
\citep{cas07}. As noted in \citet{lin08} that both cases of contamination make up only a minority of the red sequence and
blue clouds respectively, classifying different types of mergers based on their colors should be a fair approximation.

Major uncertainties in converting the pair fraction into the merger
fraction and merger rates come from the handling of the fraction of
pairs that will merge $C_{mg}$ and merger time-scale $T_{mg}$
\citep{kit08,lot08b}. The most common way of assessing merger
time-scale is computed as the "dynamical friction time"
\citep{bin87,wet09} or is estimated from the N-body/hydrodynamic
simulations of galaxy mergers \citep{con06,jiang08,lot08b}. On the
other hand, the fraction of pairs selected with projected separation
with/without the line-of-sight velocities that are
physically-associated pairs and will merge within a short time is
often obtained by correcting for chance projection
\citep{pat08,bun09}. Both $T_{mg}$ and $C_{mg}$ are usually assumed to be a
constant at a given redshift and mass bin, independent of the
environment. However, such approaches may not be adequate when
comparing the merger rate across different environments because close
pairs in dense environments are not simply isolated two-body systems,
but are also influenced by nearby galaxies, surrounding material, and
the gravitational potential from the group/cluster host halos. In
order to make a fair comparison of merger frequencies across various environments, we adopt an improved estimate of $T_{mg}$ and $C_{mg}$ as a function of environment obtained from cosmological simulations when converting the pair fraction into the fractional merger rate.

The paper is organized as follows. In \S 2, we describe our sample selection and the approach of measuring environment. In \S3, we present our results on the pair fractions for blue and
red galaxies, the computation of both $T_{mg}$ and $C_{mg}$, as well as the derived fractional merger rates for different merger categories. A discussion is given in
\S4, followed by our conclusions in \S 5. Throughout this paper we adopt the following cosmology: H$_0$ = 100$h$~\kms Mpc$^{-1}$, $\Omega_m =
0.3$ and $\Omega_{\Lambda } = 0.7$. The Hubble constant $h$ = 0.7 is adopted when calculating rest-frame
magnitudes. Unless indicated otherwise, magnitudes are given in the AB system.

\section{DATA, SAMPLE SELECTIONS, AND METHODS}
\subsection{The DEEP2 Redshift Survey}
The DEEP2 Redshift Survey (DEEP2 for short) has measured redshifts
for $\sim50,000$ galaxies at $z\sim 1$ \citep{dav03,dav07} using the
DEIMOS spectrograph \citep{fab03} on the 10-m Keck II telescope.
The survey covers four fields with Field 1 (EGS: Extended Groth
Strip) being a strip of 0.25 $\times$ 2 square degrees and Fields
2, 3 and 4 each being 0.5 $\times$ 2 square degrees. The
photometry is based on $BRI$ images taken with the 12K$\times$8K
camera on the Canada-France-Hawaii Telescope \citep{coi04}. Galaxies are selected for spectroscopy using a limit
of $R_{AB}=24.1$ mag. Except in Field 1, a two-color cut was also
applied to exclude galaxies with redshifts $z < 0.75$. A 1200
line/mm grating (R $\sim$ 5000) is used with a spectral range of
$6400 \-- 9000$ \AA, where the [OII] 3727 \AA \ doublet would be
visible at $z\sim0.7 \-- 1.4$.
The data used here contains $\sim 20,000$ galaxies with reliable redshift measurements from Fields 1, 3 and 4.
The rest-frame $B$-band magnitudes (\abm) and $U - B$ colors for DEEP2 galaxies at $0.75
< z < 0.9$ are derived in a similar way as \citet{wil06}. For galaxies with $0.9 < z <
1.2$, the rest-frame $U - B$ color is computed using the observed $R - z_{\mathrm{mega}}$ color whenever it is available, where
$z_{\mathrm{mega}}$ is the $z$-band magnitude obtained from CFHT/Megacam observations for DEEP2 Fields in 2004 and
2005 (Lin, L. et al., in preparation).

\subsection{Close Pair Sample in DEEP2}
The DEEP2 close pairs used in this work are identical to those described in \citet{lin08}. We begin with a sample of galaxies
covering $-21 <$ \mbe $< -19$ (AB mag), where \mbe is the
evolution-corrected absolute magnitude, defined as \abm + $Qz$. The
value of $Q$ is chosen to be $1.3$ in order to select galaxies with the same range relative to the
$L^{*}$ of the evolving luminosity function \citep{fab07}. Kinematic
close pairs are then identified such that their projected separations
(\dis) satisfy 10 \kpc $\leq$ \dis $\leq$ $r_{max}$ (physical length)
and rest-frame relative velocities (\vel) are less than 500 \kms\citep{pat00,lin04}. In this work, we identify the pairs using $r_{max}$ = 50 \kpc in order to have sufficient sample when dividing pairs into several different environment bins.

Galaxies are further divided into the blue cloud and red sequence using the rest-frame magnitude dependent cut for
DEEP2 (in AB magnitudes):
\begin{equation}\label{color}
U - B = -0.032(M_{B} + 21.62) + 1.035.
\end{equation}

Blue-blue pairs, red-red pairs, and blue-red
pairs are classified according to the rest-frame $U - B$ color combination of the
galaxies comprising the pair, representing the candidates of `wet',
`dry', and `mixed' mergers \citep{lin08}. In total, we have 101 blue-blue pairs, 26
red-red pairs, and 52 blue-red pairs over the redshift range $0.75 < z
< 1.2$.

\subsection{Local Environment Indicator}
For each galaxy in the DEEP2 redshift sample, the local density environment is measured using the projected third-nearest-neighbor surface density ($\Sigma_{3}$). The detailed procedure is described in \citet{coo05,coo06}. Here we briefly summarize the steps of computing the overdensity \d3 used in this work. \s3 is first calculated as \s3 = 3/$(\pi D^{2}_{p,3})$, where $D^{2}_{p,3}$ corresponds to the projected distance of the third-nearest neighbor that is within the line-of-sight velocity interval of $\pm1000$ \kms. For each galaxy, we then derive the overdensity \d3 as the local-sky completeness-corrected density \s3/$w_{p}$, denoted as \s3$^{'}$, divided by the median density \s3$^{'}(z)$ at that redshift computed in bins of $\triangle z = 0.04$:
\begin{equation}\label{delta3}
1 + \delta_{3} = \Sigma_{3}^{'} / \mathrm{median}(\Sigma_{3}^{'}(z)),
\end{equation}
where $w_{p}$ is the local-sky completeness. \1d3 is thus a measure of the overdensity relative to the median density, which takes into account the variation in the redshift dependence of the sampling rate. As discussed in \citet{coo05}, \d3 is shown to be a robust environment measure for the DEEP2 sample.

\subsection{Galaxy Group Catalogs}
Although the overdensity \d3 is a good representation of local
environment, it does not carry specific  information regarding what
kind of physical environment like field, groups, and clusters it
corresponds to. A complementary way to classify the environment is to
cross reference the galaxy sample to the group/cluster
catalogs. However, performing the group/cluster finding is not always
possible, depending on the availability of redshifts, sampling rate,
and other wavelength data (e.g. X-ray). DEEP2 targeted $\sim 65\%$ of
all of the galaxies down to R=24.1, and $\sim 70\%$ of these galaxies
yield successful redshifts. Hence the overall redshift sampling rate
of DEEP2 is about 50\%, which allows identifying potential group candidates. The group catalog used in this paper is based on the version generated by \citet{ger07}, who applied the Voronoi-Delaunay Method (VDM) group finder \citep{mar02} on the DEEP2 redshift sample. For detailed discussion on the DEEP2 group catalog, see Gerke et al. (2005, 2007). It was shown that this group catalog is most sensitive to groups with modest virial masses in the range $5 \times 10^{12} < M_{vir} < 5 \times 10^{13}M_{��}$ ($200 < \sigma_{v} < 400$ \kms) \citep{coi06}. In \S3.2, we present the distributions of various types of pairs against group properties. However, due to the incompletenss of group member identifications in the DEEP2 sample, we therefore focus our final discussion on the results obtained using the local density measurements.

\subsection{The selection function and the spectroscopic weight \label{sec:weight}}
As mentioned in \S 2.4, the overall redshift sampling rate
of DEEP2 is about 50\%, which has an impact on measuring the true pair fraction and hence the merger rate. In order to recover the intrinsic number of pairs, one must consider the completeness corrections accounting for the spectroscopic selection effects. Detailed calculations and results of the DEEP2 selection functions were presented in our previous work \citep{lin08}; here we summarize main steps of these calculations. To measure the spectroscopic weight $w$ for each galaxy
in the DEEP2 survey, we compared the sample with successful redshifts
to all objects in the photometric catalog that satisfy the survey's
limiting magnitude and any photometric redshift cut. We parameterize
the selection function to be \citep{lin08,yee96,pat02}:
\begin{equation}\label{weight}
S = S_{m}\;\overline{S_{c}}\;\overline{S_{SB}}\;\overline{S_{xy}} =
S_{m}(R)\frac{S_{c}(B-R,R-I,R)}{S_{m}(R)}\frac{S_{SB}(\mu_{R},R)}{S_{m}(R)}\frac{S_{xy}}{S_{m}(R)},
\end{equation}
where $S_{m}$ is the magnitude selection function, $S_{c}$ is the
apparent color selection function, $S_{SB}$ is the surface brightness
selection function and $S_{xy}$ represents the geometric (local
density) selection function. $\overline{S_{c}}$, $\overline{S_{SB}}$,
and $\overline{S_{xy}}$ are all normalized to the magnitude selection
function, $S_{m}$. The spectroscopic weight $w$ for each galaxy is
thus $1/S$, which is derived from its apparent $R$ mag, $B-R$ and
$R-I$ colors, $R$ band surface brightness, and local galaxy density.

The magnitude selection function $S_{m}(R)$ for each galaxy is
computed as the ratio between the number of galaxies with good
redshift qualities to the total number of galaxies in the target
catalog in both cases, considering a magnitude bin of $\pm$0.25 mag
centered on the magnitude and colors of the galaxy. The color
selection function $S_{c}$($B-R,R-I,R$) is computed by counting
galaxies within $\pm$ 0.25 $R$ magnitude over a $B-R$ and $R-I$ color
range of $\pm$ 0.25 mag. Similarly, the surface brightness selection
function is defined within $\pm$0.25 mag in $\mu_{R}$ and $\pm$ 0.25
mag in $R$. The geometric selection function $S_{xy}$($xy,R$) is
similar to the magnitude selection function but computed on a
spatially-defined (i.e., localized) scale. We take the ratio between
the number of galaxies with good quality redshifts and the total
number in the targeted catalog in an area of radius 120" within a
$\pm$ 0.25 $R$-magnitude range.

Besides the selection function for each individual galaxy, we also
investigate the selection dependence on pair separation. We measure
the angular separation of all pairs in the redshift catalog (z-z
pairs) and in the target catalog (p-p pairs) respectively and then
count the number of pairs ($N_{zz}$ and $N_{pp}$) within each angular
separation bin. While counting the pairs in the redshift catalog, each
component of the pair counts is weighted by the geometric selection
function $S_{xy}$($xy$) to exclude any effect due to the variance in
the local sampling rate. The angular selection function $S_{\theta}$
is computed as the ratio between the weighted $N_{zz}$ and
$N_{pp}$. The angular weight, $w_{\theta}$, for each galaxy is hence
$1/S_{\theta}$.

\section{RESULTS}
\subsection{Environment Distribution of Blue-Blue/Red-Red/Mixed Pairs \label{sec:distri}}
\begin{figure*}
\includegraphics[angle=-270,width=17cm]{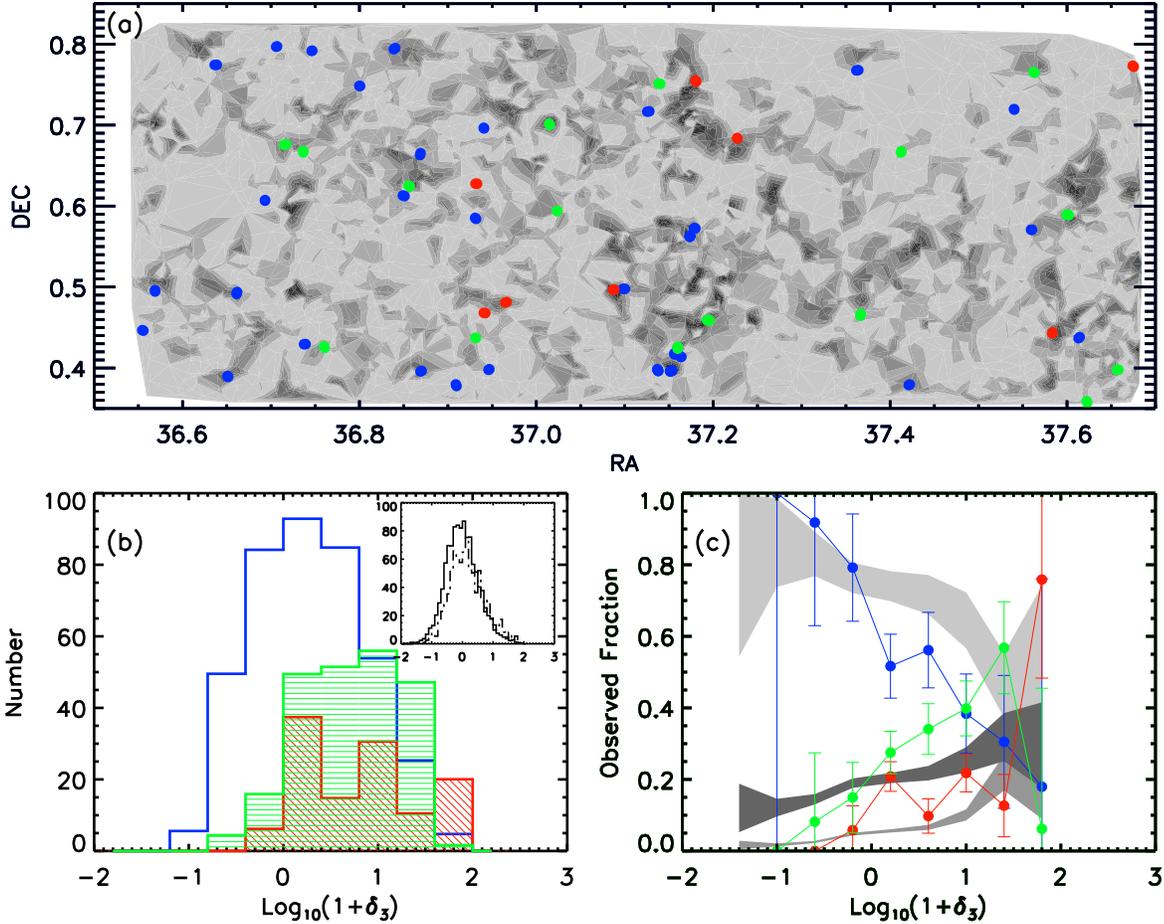}
\caption{(a) The locations of blue-blue (blue dots), red-red (red dots), and blue-red (green dots) pairs in the DEEP2 Field 4, representing candidates of wet, dry, and mixed mergers respectively in part of the DEEP2 spectroscopic sample at $0.75 < z < 1.2$. The grey contours represent the overdensity \l1d3 with 6 levels: $<$ 0, 0 - 0.3, 0.3 - 0.6, 0.6 - 0.9, 0.9 - 1.2, and $>$ 1.2 (from light to dark). (b) The distribution of local density, \1d3, for paired galaxies weighted by their spectroscopic incompleteness and angular selection functions. The paired sample is again divided into b-b (blue histogram), r-r (red histogram), and b-r (green histogram) pairs. The \1d3 distributions of the blue/red galaxies of the full sample are also shown as black solid/dash-dotted lines for comparison (the numbers of blue and red galaxies have been reduced by a factor of 12 and 4 respectively). It is clearly seen that mixed and dry mergers occur in denser environments than wet mergers do. (c) The relative fraction of b-b (blue symbols), r-r (red symbols), and b-r (green symbols) pairs as a function of \1d3. The error bars represent Poisson errors. The 1-$\sigma$ predictions based on the observed color-density relation are shown as grey areas for comparison (light to dark: wet, dry, and mixed pairs). Our comparison shows that the observed relative fraction of the three types of pairs can be explained by the change of red-galaxy fraction across different local densities. In overdense regions, however, there exists a 2-$\sigma$ difference between the observed pair fractions and the predictions following the color-density relation, which indicates that the clustering of blue and red galaxies at small scales may be different in those environments.
\label{d3distri}}
\end{figure*}

Fig. \ref{d3distri}(a) shows the projected positions of wet, dry, and
mixed pairs found in one of the DEEP2 fields (Field 4), overlaid with
contours tracing the mean density along the line-of-sight. Visually it reveals that blue-blue pairs appear in all kinds of environments, from low to high density regions. On the other hand, red-red pairs and blue-red pairs tend to lie in denser environments. Quantitative comparisons between the local environment of the three
types of pairs, after correcting for the spectroscopic incompleteness,
are shown in Fig. \ref{d3distri}(b). We first note that all blue-blue, red-red, and blue-red pairs have median local density greater than the average environment (\l1d3 $\sim 0$). This is expected: paired
galaxies, by definition, have a close companion nearby and thus their
separation from the 3$^{rd}$-nearest neighbor will be smaller (and hence denser) on average by construction. The most
interesting result of Fig. \ref{d3distri}(b) lies in the difference in
the density distribution among blue-blue, red-red, and blue-red
pairs. While blue-blue pairs favor median-density environments, red-red and blue-red pairs are preferentially located in overdense regions.

To better understand whether such density distribution of
pairs is related to the color-density relation, i.e., the change of
the fraction of red galaxies across different local densities \citep{hog04,coo06,cuc06},
we perform the following analysis: for a given local density, we
compute the red-galaxy fraction corrected by the spectroscopic
incompleteness, and then derive the predicted relative fraction among
wet, dry, and mixed pairs assuming that the red and blue galaxies are
randomly distributed. As illustrated in Fig. \ref{d3distri}(c), the
increased fractions of dry and mixed pairs with respect to the local
density follow a similar trend as expected from the color-density
relation. However, we find an excess of dry and mixed pairs toward
overdense regions at a $\sim$ 2-$\sigma$ level compared to the above
expectation, indicating that the red and blue galaxies are not
uniformly distributed and that there exists a clustering effect at very small
scales in those overdense environments.

.

\begin{figure*}
\includegraphics[angle=-270,width=17cm]{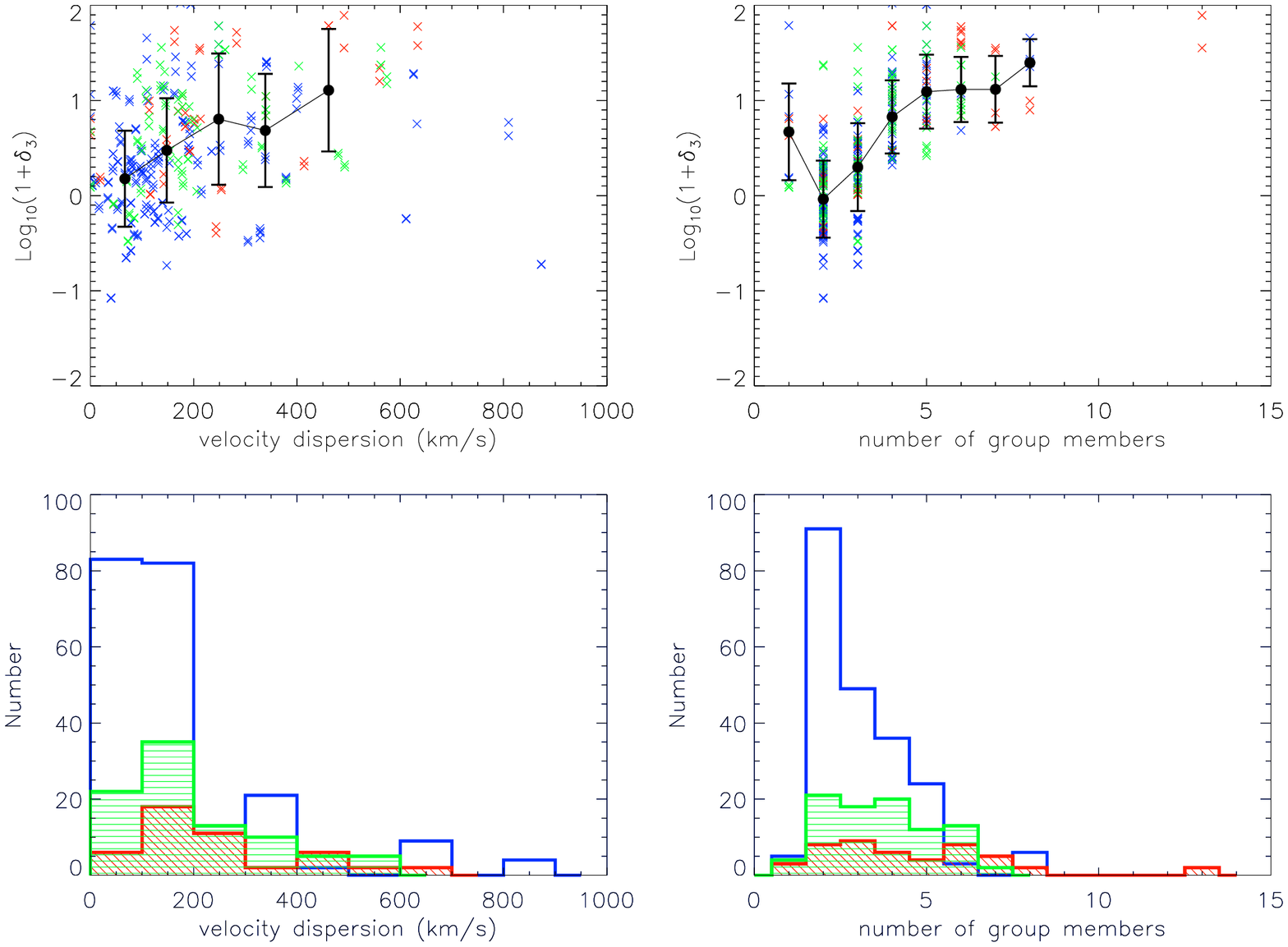}
\caption{The two upper panels plot the overdensity against velocity dispersion (left) and the number of group members (right) for paired galaxies. The black symbols and associated error bars indicate the median value with the root-mean-squared of the scatter in each bin. The two lower panels display the histograms of velocity dispersion of groups and the number of group members for groups that host paired galaxies. Blue, green, and red colors denote for blue-blue, red-red, and blue-red pairs respectively.
\label{group}}
\end{figure*}

The difference in the density distribution of wet/dry/mixed mergers
suggests that the physical environment where various types of mergers
occur is essentially different. To interpret our results, we also
investigate how different mergers are populated as a function of the velocity dispersion and the number of members of galaxy groups for a subset of DEEP2 samples that have group identifications as presented in \citet{ger07}. The two upper panels of Fig. \ref{group} plot the overdensity against velocity dispersion (left) and the number of group members (right) for paired galaxies. As expected, there is a clear trend that the local density of galaxies belonging to groups with greater velocity dispersion or group members is on average larger. This illustrates that the local environment measure \d3 used in this work in general correlates well with physical environments (field, groups, clusters). As shown in the two bottom panels of Fig. \ref{group}, the fraction of red-red and blue-red pairs in groups with greater velocity dispersion or more group members is significantly higher than that of blue-blue pairs. More specifically, the majority of blue-blue pairs are found in field-like environments while red-red and blue-red pairs tend to be found in group and/or cluster-like environments.

It is worth pointing out that there are $\sim$ 11\% of the pair sample whose two members do not belong to the same group. This can happen when a single group is split into two or more smaller groups by the group finder, or a group is not properly identified owing to the incompleteness of the spectroscopic sample. As a result, there are some pairs that are identified as 'field galaxy' based on the group finder. These are the paired galaxies assigned to have one group member as shown in the lower-right panel of Fig. \ref{group}. Such an effect makes the group results rather hard to interpret compared to the local density results, which are relatively insensitive to the spectroscopic incompleteness. We therefore focus on the discussion of the environment effects based on the results using the local density in the rest part of this paper.

\subsection{The Environment dependence of the Pair Fraction \label{pairfraction}}
The above analysis on the environment distributions of wet/dry/mixed merging galaxy pairs
provides insight into which environments play host to most galaxy
mergers. However, this is different from asking in which environments
galaxy mergers are more likely to occur. In this section, we investigate the latter issue by studying the relative frequency of galaxy interactions across different environments, i.e., to count the paired galaxies relative to the parent sample as a function of environment. We note that the pair fraction does not necessarily correspond to the merger fraction because not every kinematic pair defined observationally will
eventually merge into one system. Such phenomenon is in particular more frequent in dense environments due to chance projection as well as many-body interactions. We will address this issue in \S \ref{CmgTmg} and \S \ref{mergerfraction}.

The method we adopt to compute the pair fraction, \Nc, is the same as described in our previous work \citep[see \S 3.1 of ][]{lin08}, except that we
further bin the sample by the local densities. The pair fraction \Nc is defined as the average number of companions per galaxy:
\begin{equation}\label{Nc}
N_{c}=\frac{\sum^{N_{tot}}_{i=1}\sum_{j}w_{j}w(\theta)_{ij}}{N_{tot}},
\end{equation}
where $N_{tot}$ is the total number of galaxies within the chosen absolute magnitude range, $w_{j}$ is the
spectroscopic weight for the $j$th companion belonging to the $i$th galaxy, and $w(\theta)_{ij}$ is the angular
selection weight for each pair as described in \S 2.5. The averaged value of the overall spectroscopic weight $w$ of our paired galaxies is about 2.1 and that of the angular selection weight is about 1.2 in the redshift range of $0.75 < z < 1.2$. A correction factor of
2.4 in addition to the usual spectroscopic and angular separation corrections is also applied for each red companion at $z > 1$ to account for the missing faint red galaxies in the high-redshift DEEP2 sample \citep{lin08}.
Four types of pair fraction are measured here: a) \Nc from all pairs regardless of colors;
b) the average number of blue companions per blue galaxy \Ncb; c) the average number of red companions per red
galaxy  \Ncr; d) the average number of companions of galaxies with opposite color to that of the primary galaxies
\Ncm. Note that b) and c) are equivalent to the pair fraction within
the blue cloud and red sequence, respectively. Here we divide the
environment into three regimes: underdense environment, intermediate
environment, and overdense environment. Naively one would think that
the pair fraction should increase with the local density. In fact,
this is not necessarily true for the blue pair fraction (\Ncb), red
pair fraction (\Ncr), or mixed pair fraction (\Ncm) individually. One
must keep in mind that the overdensity is computed using all galaxies
of all colors and not segregated by galaxy colors. Therefore how the \Ncb, \Ncr, and \Ncm vary against environment depends on the relative red and blue fraction at a given environment, as discussed in \S3.1

\begin{figure*}
\includegraphics[angle=-90,width=17cm]{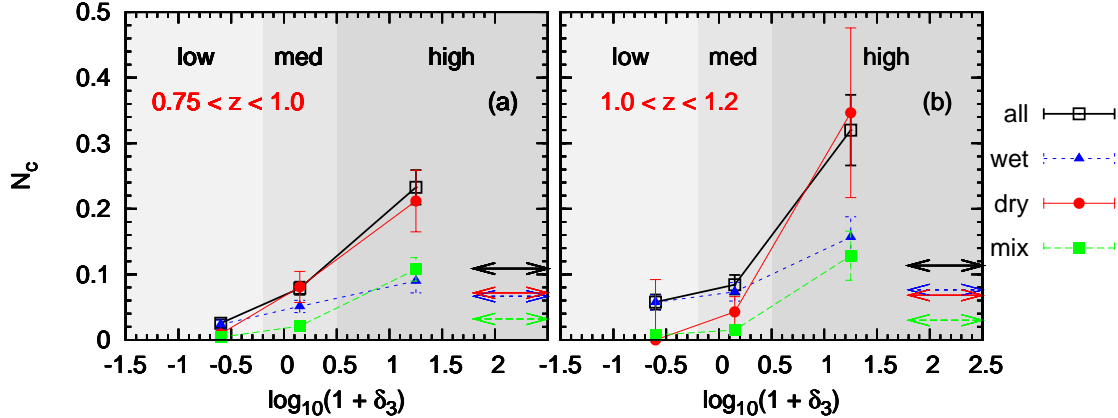}
\caption{The pair fraction as a function of overdensity \1d3. Here we show the results for four types of pairs: blue-blue pairs (blue triangles), red-red pairs (red solid circles), blue-red pairs (green solid squares), and all pairs regardless their colors (black open squares). The denominators used for computing the above four quantities are the numbers of blue, red, blue+red, and blue+red galaxies respectively. The error bars shown in the plot are calculated by bootstrapping. We set \Nc = 0 and put the errors to be the Poisson errors for 5 objects when no pair is found at a given environment. The horizontal arrows appeared in the right-lower corner of each panel indicate the global pair fraction in the same redshift range taken from Lin et al. (2008) computed without separating data into different environment bins. There exists clear environment dependence of the pair fraction in the considered two redshift bins, being higher at denser environments.
\label{nc-d3}}
\end{figure*}

Fig. \ref{nc-d3} displays the pair fraction as a function of overdensity \1d3 in log space for two redshift bins $0.75 < z < 1.0$ and $1.0 < z < 1.2$.
The rapid rise of \Nc with increasing density is similar for all four types of pairs for the two redshift bins considered. However, the relative companion rate among wet/dry/mixed pairs changes across different environments. In underdense regions, the blue companion rate for blue galaxies (blue points) is in general higher than the red companion rate for red galaxies (red points). On the other hand, the opposite holds in overdense environments. This suggests that while the overall companion rate is enhanced in dense environments, the level of enhancement depends on the types of galaxies. In order to visualize the enhancement of the companion rate as a function of environment, the global values of \Nc at a given redshift obtained by \citet{lin08} are also shown as horizontal arrows in each panel of Fig. \ref{nc-d3}. Quantitatively speaking, the companion rate is about 1.3-5 times greater in high-density regions compared to the global rate, depending on the type of pair.

\subsection{Estimates of $C_{mg}$ and $T_{mg}$ as a Function of the Local Environment \label{CmgTmg}}
\begin{figure}
\includegraphics[angle=-0,width=9.0cm]{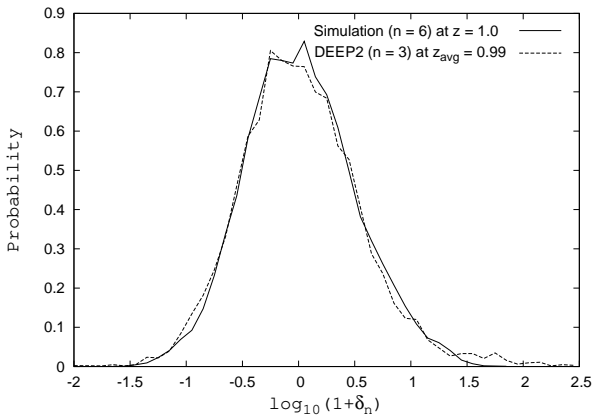}
\caption{The overdensity \onedn distributions of the mock galaxy catalog (solid curve) and of the DEEP2 sample (dashed curve) at $z \sim 1$. The projected distance from the 3$^{rd}$-nearest neighbor is adopted to compute the overdensity for the DEEP2 sample; that of the 6$^{th}$-nearest neighbor is used for overdensity measurements in the mock galaxy catalog. The different choices of $n^{th}$-nearest neighbors between two samples are due to the spectroscopic incompleteness of the DEEP2 survey. Empirically we found that the adoption of $n = 6$ in the simulated galaxy catalog best reproduces the overdensity \1d3 distribution function of the observed DEEP2 galaxies.
\label{fig:dndistri}} \end{figure}

As mentioned in \S \ref{pairfraction}, the pair fraction is not equal to the galaxy merger fraction unless all pairs are merging systems. It is possible that the pair
sample is subject to contamination from interlopers owing to the
difficulty of disentangling the Hubble expansion and the galaxy peculiar velocity.
In order to take into account any possible environment effect on the merger time-scale ($T_{mg}$) and the fraction of merger in pairs ($C_{mg}$), we construct a mock galaxy catalog based on the dark matter halos and subhalos taken from a cosmological N-body simulation, and trace merger
histories of halo-halo pairs to examine the dependence of $C_{mg}$ and $T_{mg}$ on the local density. A full description will be presented in a forthcoming paper (Jian, H.-Y. et al., in preparation). Here we briefly describe the techniques that are used to study this problem and present the most relevant results. We make use of the cosmological N-body simulations and the dark matter halos as presented in \citet{jia08}. The simulation used here has been evolved in the concordance flat $\Lambda$CDM model: $\Omega_{m}$ = 0.3, $\Omega_{\Lambda}$ = 0.7, and $\Omega_{b}$ = 0.05. It contains 512$^{3}$
pure dark matter particles in a 100 $h^{-1}$ $Mpc$ box on a side. The resulting mass of a dark matter particle is $m_{dm}$ = 6.188 $\times$ 10$^{8}$$h^{-1}$$M_{\odot}$. The distinct halos and substructures (subhalos) are identified using a variant version of Hierarchical Friends-of-Friends Algorithm \citep[HFOF]{kly99}, with the minimum particle number of 30. The close halo-halo pairs are selected in a way that they satisfy the observational criteria in both projected separation and in line-of-sight velocity difference to mimic the observed pairs. The halos in pairs can be either distinct halos (no smaller substructures contained or not hosted by a larger halo) or subhalos. For each halo, we compute its local density using the separation from its nearest $n^{th}$ neighbor that is above certain mass cut $M_{min}$. Both $n$ and $M_{min}$ are determined so as to match the median distance to the $3^{rd}$-nearest neighbor in the DEEP2 sample. Empirically, we found that $n = 6$ in the simulations traces the same comoving scale as $n=3$ does in the observed data
set. We note that the overall DEEP2 redshift completeness is $\sim 50\%$, which means that the observed $3^{rd}$-nearest neighbor roughly corresponds to the `true' $6^{th}$- or $7^{th}$-nearest neighbor. Therefore, our choice of $n = 6$ in the simulations should be a reasonable approach. In the rest of this paper, we refer \dn to the overdensity measured using the $6^{th}$ nearest neighbor for halos. The resulting \onedn distribution of halos is very similar to the \1d3 distribution of observed DEEP2 galaxies, as demonstrated in Fig. \ref{fig:dndistri}.

\begin{figure*}
\includegraphics[angle=-0,width=16.0cm]{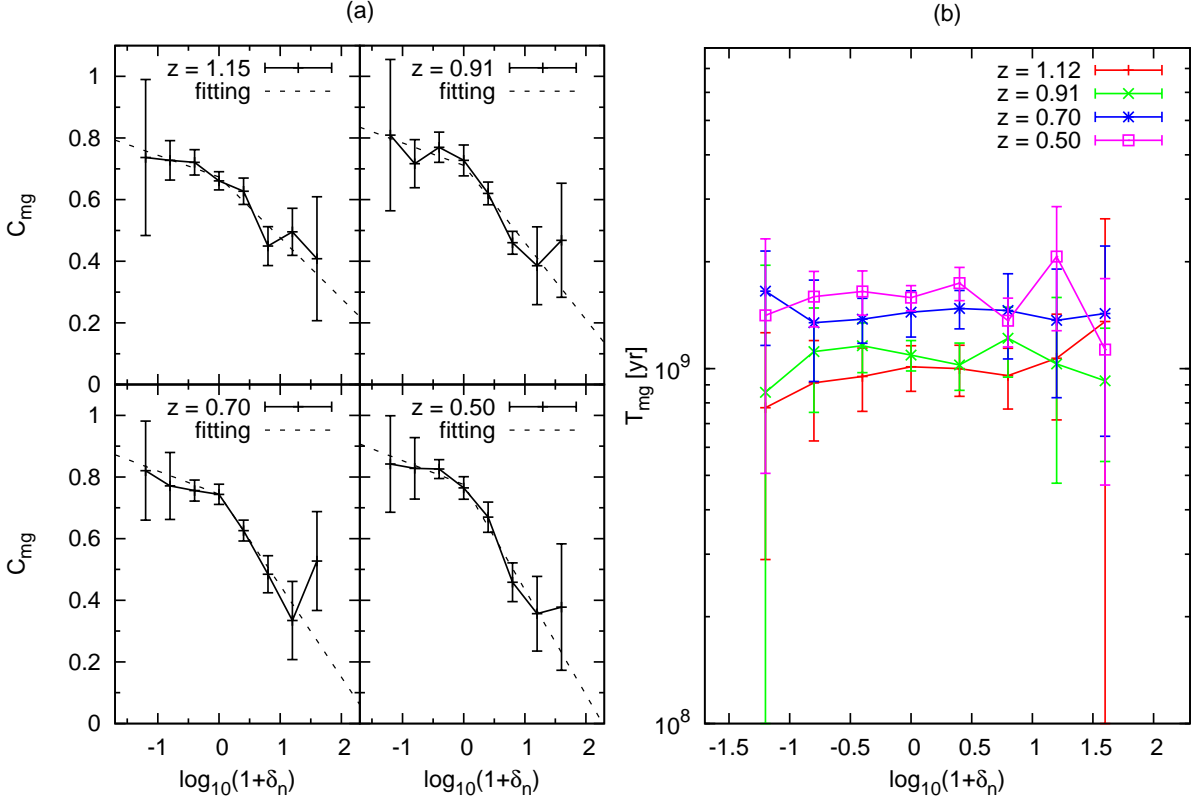}
\caption{(a) $C_{mg}$ (fraction of kinematic pairs to be merged) as a function of overdensity \onedn, determined using the mock galaxy catalog constructed from the N-body simulations. The dashed lines represent the best-fitting formula of the data points. Over the entire redshift range we have probed, $C_{mg}$ is a strong function of local environment. This is largely due to the stronger projection effects in overdense regions than that in underdense regions. The fitting formula of $C_{mg}$ is given in Eq. \ref{cmg}. (b) $T_{mg}$ (the merging time-scale) as a function of overdensity \onedn for those pairs that will eventually merge. The error bars represent the root-mean-squared of the scatter in each bin.
\label{ct}} \end{figure*}

For each halo-halo pair identified with the criteria \dis $\leq$ $50$ \kpc and \vel $\leq$ 500 \kms, we trace their most-bounded 10 particles identified at a given epoch in the next adjacent few redshift frames. If 60\% of these most-bounded 10 particles from the two pair components can be found in a single halo in the sequential frame, the pairs are  then called to be merged. $C_{mg}$ is thus computed as the fraction of pairs that will merge into a single halo. Among those merged halos, we record the time-scale $T_{mg}$ over which the halo-halo pairs merge.

Fig. \ref{ct} displays $C_{mg}$ and $T_{mg}$ as a function of local density. It is interesting that while $T_{mg}$ varies little with the local density, $C_{mg}$ is a strong function of \onedn, being smaller in higher density regions. This suggests that different environments have a strong impact on determining whether the close pairs will merge or not, but have little influence on the merger time-scale if those pairs are going to merge. When we analyze those halo-halo pairs that do not merge, we find that the majority of these pairs are actually widely separated in 3-D space and have large differences in 3-D velocity, and such projection effects are more pronounced for pairs in overdense environments.
In the remainder of the cases, one component of the halo pairs may be tidally stripped and fails to be identified as a halo if they do not satisfy the virial condition in the next snapshot \citep{jia08}. In such cases, if the most-bounded particles do not belong to any halo, we then stop tracing their histories and count them as non-merging cases. We caution that this might underestimate $C_{mg}$  in a way that the galaxy component may still survive temporarily until it merges with its companions, even though the dark matter of its hosting halos is stripped. However, if this is true, they will contribute to the tail of the $T_{mg}$ distribution and hence shift $T_{mg}$ toward a higher value. Because the merger rate is proportional to $C_{mg}$ and inversely proportional to $T_{mg}$, such an effect will be roughly canceled out. To model the environment dependence of $C_{mg}$, we fit the curves in panel (a) of Fig. \ref{ct} by two lines:

\begin{equation}\label{cmg}
C_{mg} = \left\{
    \begin{array}{ll}
	\mbox{$(0.01z - 0.08)x - (0.16z - 0.85)$,} & \mbox{if $x < 0$}\\
        \mbox{$(0.22z - 0.45)x - (0.16z - 0.85)$,} & \mbox{if $x \geq 0$},
    \end{array}
\right.
\end{equation}
where $x$ = \pdn.

In the simulations, the value of $T_{mg}$ for pairs with \dis $<$ 50
\kpc is approximately 1 Gyr at $z \sim 1$, which is almost twice the
typical value of $\sim$ 0.5 Gyr adopted in previous
studies \citep{lin04,lin08} that used a more stringent criterium \dis $<$ 30 \kpc. The longer time-scale for such wider pairs has also been suggested in earlier works by \citet{lot08b} who studied the merger time-scale using N-body/hydrodynamical simulations. We notice that $T_{mg}$ increases slightly when going to lower redshifts. Because the simulations were stored at discrete epochs, the value of $T_{mg}$ can only be estimated by summing the time interval of several adjacent frames until the last frame in which the halos are identified as merged. In this way, $T_{mg}$ is likely to be overestimated. However, since the time interval is typically $\sim$ 200 Myr at $z \sim 1$, which is much smaller than the typical time-scales normally found for pairs with \dis $< 50$ \kpc \citep{lot08b}, we believe such uncertainty is negligible. Such an effect, on the other hand, becomes more apparent at lower redshift as the time interval between two adjacent redshift frames of the simulations rises to $400$ Myr at $z \sim 0.4$. This might explain the trend of increasing $T_{mg}$ with redshift. In this work, we adopt $T_{mg} = 1$ Gyr in all environment and $C_{mg}$ derived with Eq. \ref{cmg} when converting the pair fraction into the fractional merger rate as presented in the next section.

Before proceeding to compute the merger rate inferred from the pair fraction, it is worth discussing how our derived $T_{mg}$ and $C_{mg}$ are compared to previous works by other groups, in order to assess possible systematic errors in our estimates in the fractional merger rate. As we will see in \S \ref{mergerfraction}, the fractional merger rate is proportional to $C_{mg}$/$T_{mg}$, here we use the ratio $C_{mg}$/$T_{mg}$ as a comparison quantity. In our case, the typical value of $C_{mg}$ is approximately 0.7 and $T_{mg}$ is $\sim 1$ Gyr averaged in all kind of environments, leading to $C_{mg}$/$T_{mg}$ = 0.7. A recent study by \citet{kit08} use the Millennium Simulation \citep{spr05} to determine the averaged merger time-scale $T_{mg}$ as a function of stellar mass and redshift of close pairs selected with various selection criteria \citep[see Eq. (10) and (11) in ][]{kit08}. In their analysis, every pair will eventually merger; in other words, the effect of $C_{mg}$ is absorbed into the quantity of $T_{mg}$ (i.e., equivalent to set $C_{mg} = 1$). If adopting their Eq. (10) with $h$ = 0.7, \dis $<$ 50 \kpc and the stellar mass \sm $\sim 3\times10^{10}M_{\odot}$, which is the typical stellar masses in our pair sample, we get $T_{mg} \sim$ 2.7 Gyr. The ratio of $C_{mg}$/$T_{mg}$ is thus 0.37, which is about half of our value of 0.7. This leads to a potential uncertainty by as large as a factor of two in the estimates of the fractional merger rate, depending on the adopted modeling of $C_{mg}$ and $T_{mg}$.

\subsection{The Fractional Merger Rate $f_{mg}$ as a Function of Environment \label{mergerfraction}}
In this section, we present our results on the fractional merger rate,
defined as the fraction of galaxies in the range of $-21 < $\mbe $<
-19$ that merge per Gyr with another galaxy such that the luminosity
ratio of the pair is between 4:1 and 1:4. This quantity can be derived
from the pair fraction computed in \S \ref{pairfraction} with the
knowledge of $C_{mg}$ and $T_{mg}$ we have obtained in \S
\ref{CmgTmg}, but keeping in mind that the pair fraction in \S 3.2 is computed using pairs drawn from within a luminosity range of two magnitudes. Some true companions may
fall outside the absolute magnitude range of our sample, while some selected companions have luminosity ratios
outside the range of 4:1 to 1:4. To account for both of these effects, we use the following equation to convert the
pair fraction into the fractional merger rate $f_{mg}$:
\begin{equation}\label{f_mg}
f_{mg}=(1+G)\times C_{mg}N_{c}(z)T^{-1}_{mg},
\end{equation}
where $G$ is the correction
factor that accounts for the selection effect of companions due to the restricted luminosity range (see Lin et al. 2008 for the detailed computation of $G$). It is worth noting that the factor $(1+G)$ in Eq. \ref{f_mg} is different from $(0.5+G)$ that is shown in the Eq. (5) of \citet{lin08} due to different definitions between the merger rate \citep{lin08} and the fractional merger rate adopted in this work.

\begin{figure*}
\includegraphics[angle=-90,width=17cm]{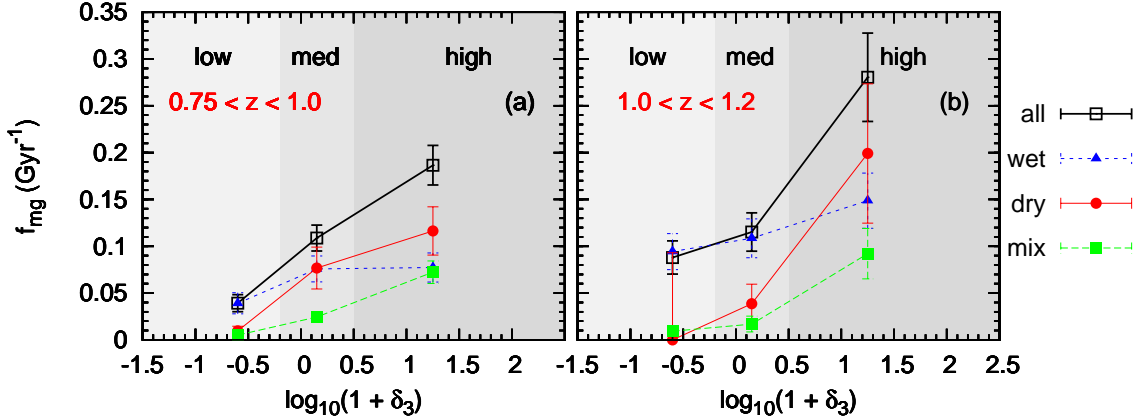}
\caption{The fractional merger rate, $f_{mg}$, as a function of overdensity \1d3. Here we show the results for four types of mergers: wet mergers (blue triangles), dry mergers (red solid circles), mixed mergers (green solid squares), and all pairs regardless of their colors (black open squares). The above four quantities have been normalized to the numbers of blue, red, blue+red, and blue+red galaxies respectively. The error bars are calculated by bootstrapping. We set $f_{mg}$ = 0 and put the errors to be the Poisson errors for 5 objects when no pair is found at a given environment. While the fractional wet merger rate shows weak dependence on local density, the fractional rate of dry and mixed mergers strongly depends on the overdensity in the redshift range probed.
\label{fmg}} \end{figure*}

In Fig. \ref{fmg}, we show the fraction merger rate, $f_{mg}$, as a function of overdensity for wet (blue points), dry (red points), mixed (green points), and all (black points) mergers. Those values are also listed in Table \ref{tab:fmg}. Owing to the decreasing $C_{mg}$ with overdensity, the increase of $f_{mg}$ with respect to the overdensity is not as steep as \Nc.
However, we still find that the fractional merger rate in the overdense regions is, on
average, 3-4 times greater than that in the underdense regions for all
mergers regardless of their types, shown as black symbols in Fig. \ref{fmg} (also see Table \ref{tab:fmg}). Such
enhancement in dense environments is in broad agreement with recent
theoretical work by \citet{fak09} who measured the merger rates of
friends-of-friends (FOF) identified mock matter halos in the Millennium
simulation \citep{spr05} as a function of local mass density. When
dividing the merger sample into subcategories (wet, dry, and mixed
mergers) we find a significant enhancement of the frequency of dry and mixed mergers between under- and overdense environments. In
contrast, there is only a weak environment dependence between density extremes for wet mergers
(Fig. \ref{fmg}). This implies that the group-like and cluster-like environment are preferred environments for dry and mixed mergers to take place.

At $z \sim 1$, the fractional dry
merger rate in high-density regions is found to be 16$\pm$4\% (Table~1), which
is about 3 times larger than the global fractional dry merger rate
derived from earlier studies \citep{lin08,bun09} regardless of their
environments. An enhancement of dry mergers in overdense environments
is also observed in the local universe. Using data drawn from the
Sloan Digital Sky Survey (SDSS), \citet{mci08}
find that the frequency of mergers between luminous red galaxies
(LRGs) is higher in groups and clusters compared to that of overall
population of LRGs by a factor of $2-9$ at $z < 0.12$. Our work
similarly suggests that a greater probability for dry mergers
in high density environments was already in place by at least
$z \sim 1$. If we assume an average stellar mass ratio of 1:2 in our
dry merger sample, a constant fractional dry merger rate in dense
environments at $0 < z < 1$, and that all stellar mass involved in
each merger is deposited into the final merger remnant, then we estimate
that on average every local massive red-sequence galaxy in a dense
environment is assembled through $0.16\pm0.04 (merger/Gyr) \times 7.7(Gyrs)
\sim 1.2\pm0.3$ major dry mergers, leading to $\sim$ (38$\pm$10)\% ( = $1.2 \times
0.5 /(1 + 1.2 \times 0.5))$ mass accretion since $z \sim 1$.

\begin{deluxetable*}{lcccccccccccc}
\tabletypesize{\scriptsize}
\tablewidth{0pt}
\tablecaption{The fractional merger rate ($f_{mg}$) as a function of different environment \label{tab:fmg}}
\tablehead{
    \colhead{Merger Types} &
    \colhead{$\overline{z}$} &
    \colhead{$N_{c}^{a}$} &
    \colhead{$N_{c}^{b}$} &
    \colhead{$N_{c}^{c}$} &
    \colhead{$G$} &
    \colhead{$C_{mg}^{a}$} &
    \colhead{$C_{mg}^{b}$} &
    \colhead{$C_{mg}^{c}$} &
    \colhead{$T_{mg}^{a}$} &
    \colhead{$f_{mg}^{a}$} &
    \colhead{$f_{mg}^{b}$} &
    \colhead{$f_{mg}^{c}$} \\
    \colhead{}   &
    \colhead{}   &
    \colhead{}   &
    \colhead{}   &
    \colhead{}   &
    \colhead{}   &
    \colhead{}   &
    \colhead{}   &
    \colhead{}   &
    \colhead{(Gyr)}   &
    \colhead{(Gyr$^{-1}$)}   &
    \colhead{(Gyr$^{-1}$)}   &
    \colhead{(Gyr$^{-1}$)}
}

\startdata
All     &0.88   &0.025$\pm$0.006   &0.079$\pm$0.010   &0.233$\pm$0.026   &1.04   &0.76   &0.68   &0.39   &1.0   &0.039$\pm$0.009    &0.109$\pm$0.014    &0.187$\pm$0.021 \\
        &1.08   &0.058$\pm$0.012   &0.084$\pm$0.015   &0.032$\pm$0.054   &1.10   &0.73   &0.65   &0.42   &1.0   &0.088$\pm$0.018    &0.115$\pm$0.020    &0.281$\pm$0.047 \\
Wet     &0.88   &0.024$\pm$0.007   &0.051$\pm$0.009   &0.090$\pm$0.018   &1.19   &0.76   &0.68   &0.39   &1.0   &0.039$\pm$0.011    &0.076$\pm$0.014    &0.077$\pm$0.016 \\
        &1.08   &0.057$\pm$0.012   &0.073$\pm$0.014   &0.157$\pm$0.031   &1.27   &0.73   &0.65   &0.42   &1.0   &0.094$\pm$0.019    &0.109$\pm$0.021    &0.149$\pm$0.030 \\
Dry     &0.88   &0.009$\pm$0.005   &0.081$\pm$0.024   &0.212$\pm$0.047   &0.40   &0.76   &0.68   &0.39   &1.0   &0.010$\pm$0.005    &0.077$\pm$0.022    &0.116$\pm$0.026 \\
        &1.08   &0.000$\pm$0.092   &0.043$\pm$0.023   &0.347$\pm$0.129   &0.38   &0.73   &0.65   &0.42   &1.0   &0.000$\pm$0.093    &0.039$\pm$0.021    &0.199$\pm$0.074 \\
Mixed   &0.88   &0.004$\pm$0.002   &0.021$\pm$0.005   &0.108$\pm$0.018   &0.71   &0.76   &0.68   &0.39   &1.0   &0.005$\pm$0.003    &0.025$\pm$0.005    &0.072$\pm$0.012 \\
        &1.08   &0.007$\pm$0.004   &0.015$\pm$0.008   &0.128$\pm$0.038   &0.72   &0.73   &0.65   &0.42   &1.0   &0.009$\pm$0.005    &0.017$\pm$0.008    &0.092$\pm$0.027
\enddata

\tablenotetext{a}{$low$ density (-1.0 $<$ \l1d3 $<$ -0.2)}
\tablenotetext{b}{$median$ density (-0.2 $<$ \l1d3 $<$ 0.5)}
\tablenotetext{c}{$high$ density (0.5 $<$ \l1d3 $<$ 2.0)}
\end{deluxetable*}

\section{DISCUSSION}
\subsection{Comparison of the Environment Dependence of Merger Rates Between Observations and Simulations}
In this subsection, we discuss how our results are compared to
previous theoretical predictions on the environment dependence of the
merger rate of dark matter halos. There have been several attempts to
investigate the relation between halo merger rates and underlying
environments either using N-body simulations \citep{fak09,hes09} or
based on the Monte-Carlo merger trees that are constructed with the
extended Press-Schechter (EPS) and excursion set models
\citep{kau00}. Using the Millennium simulation \citep{spr05},
\citet{fak09} measured the merger rate of dark matter halos as a
function of the local mass density within a sphere of several Mpc
using a friends-of-friends (FOF) algorithm. They found a strong
dependence of specific halo merger rates on the environment, being
greater in the densest regions than in voids by a factor of $\sim
2.5$. The level of enhancement of the specific halo merger rates in
dense regions is in broad agreement with what we measure for observed
galaxies. Very recent work by \citet{hes09} has also explored similar
issues but for subhalos extracted from the Millennium simulations. In
contrast, they find that in group environments, the subhalos are often
tidally stripped and hence the chance of subhalo-subhalo mergers is
low. As a consequence, the specific halo merger rate in groups is
normally suppressed, which seems to be in contradiction to our finding
that the fractional merger rate is enhanced in overdense environments.

However, we caution that direct comparisons between our results and
simulations could be limited by several factors. For example, the
studies by \citet{fak09} utilize the merger trees constructed from
distinct FOF halos which do not correspond as well to observed galaxies
as do the subhalos (substructures). Therefore
translating their simulation results of halo merger rates into the
actual merger rate of galaxies is not straightforward. On the other
hand, the halo environment adopted in \citet{hes09} is based on the
size/mass of the groups, characterized by the maximum of their
rotation velocity curve, $V_{max}$. As shown in Fig. \ref{group},
there is spread in \1d3 for a given number of group members, and vice
versa, despite that in general the local density increases with the
global environment (velocity dispersion, the number of group members,
etc.). This suggests that at a given high local density in our sample,
there are contributions from both the 'dense' field-like environments,
as well as group-like and possibly even cluster-like
environments. Therefore more adequate comparisons shall await larger
surveys which sample mergers in various scales of galaxy groups and
clusters.

\begin{figure*}
\includegraphics[angle=-270,width=17cm]{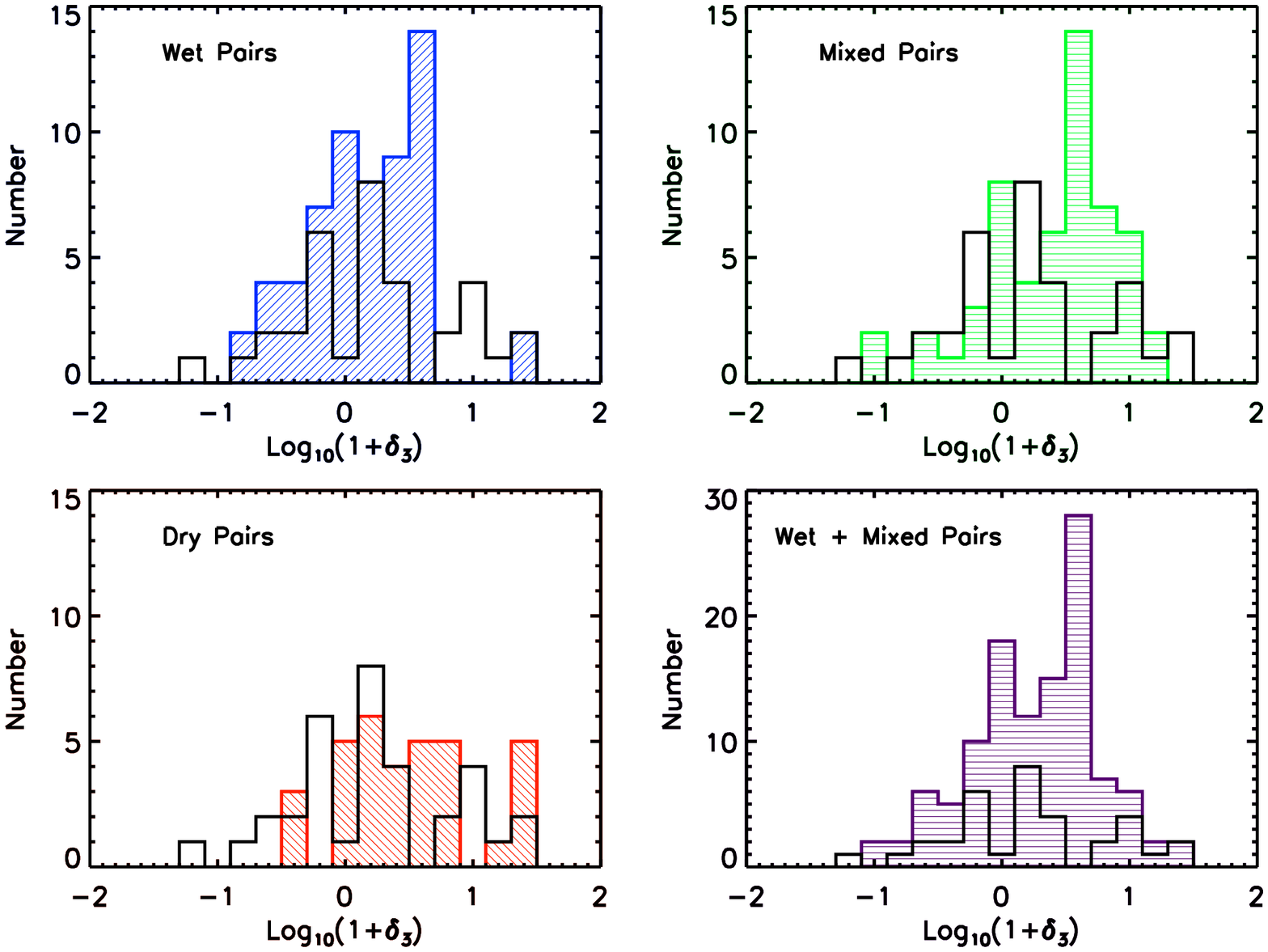}
\caption{Comparisons of the overdensity \1d3 distribution of K+A galaxies with candidates of wet (upper left panel), dry (lower-left panel), mixed (upper-right panel), and wet+mixed mergers (lower-right) respectively. The distributions of K+A galaxies are presented as black solid histograms while those of merger candidates are shown as color shaded areas.\label{KA}
\label{KA}} \end{figure*}

\subsection{Are K+A Galaxies Formed Through Major Mergers? \label{sec:KA}}
There have been several mechanisms proposed to quench the star formation in galaxies and lead to the formation of K+A galaxies. These mechanisms include galaxy-galaxy mergers \citep{mih94}, ram-pressure stripping \citep{gun72}, high speed galaxy encounters \citep[galaxy harassment;][]{moo96}, and 'strangulations' in which the warm and hot gas is removed \citep{lar80,bal00}. Except for galaxy mergers, many of those are strongly associated with the cluster environment. Several environment studies of poststarbursts  have found  a higher fraction of K+A galaxies in clusters than in the field \citep[e.g.][]{tra03,tra04,pog99}. In contrast, other studies using large low-redshift samples have suggested that poststarbursts are preferentially found in the low density region \citep{bal05,got05,hog06}. Recently, \citet{yan09} studied the environment distribution of 74 K+A galaxies found at $z \sim 0.8$ in the DEEP2 redshift survey. They found that at this redshift range, there is very little environment dependence of the K+A fraction. Putting all these results together, it is suggested that the galaxy merger, which is not a cluster-specific mechanism, is potentially an important origin of K+A galaxies found in the field. One way to test this hypothesis is to compare the environment distributions of K+A samples to that of the merger samples.

As shown in Fig. \ref{fmg}, the fractional wet merger rate in the
DEEP2 sample depends weakly on the local density, similarly to the
trend seen in DEEP2 K+A galaxies (see Fig. 6 of Yan et al. 2009). On
the other hand, the mixed mergers show stronger dependence on the
environment, unlike the K+A galaxies. In order to make more careful
comparisons, we limit the K+A sample selected by \citet{yan09} with an
additional restframe magnitude cut -21.77 $<$ \mbe $<$ -19.77, which
is 2 $\times$ brighter than the pair sample, to reflect the assumption that
the K+A galaxies are products of two merging galaxies. We also apply
the redshfit cut of $0.75 < z < 0.88$ to our pair samples so
they span the same redshift range as the K+A galaxies. In
Fig. \ref{KA}, we plot the \1d3 distribution of K+A galaxies against
wet, dry, mixed, and wet+mixed pairs. Among the four types of pair
samples, the density distributions of dry pairs and mixed pairs are
distinct from that of the K+A sample, whereas wet pairs or wet+mixed
pairs show similar a density distribution to the K+A galaxies. To quantify
the significance of the environment difference or similarity between
K+A samples and mergers, we perform three non-parametric statistical
tests, the Kolmogorov-Smirnov test (K-S test), the Anderson-Darling
test \citep[A-D test;][]{and54,pet76,sin88}, and the Mann-Whitney-Wilcoxon test \citep[MWW test;][]{man47} as done
in \citet{yan09}. The results are presented in Table
\ref{tab:kstest}. We find that the p-values derived from the K-S,
A-D, and MWW tests for the K+A sample against dry and mixed mergers
alone are all very close to the rejection threshold 0.05.
On the other hand, the p-values for the wet vs. K+A set and the
wet+mixed vs. K+A set are well beyond the threshold 0.05. Based on
these results, we conclude that the environment distributions of K+A
galaxies and of wet or wet+mixed mergers are
indistinguishable. Nevertheless, we caution that such analysis is
possibly limited by the small numbers of K+A and pair samples.

The idea that K+A galaxies could be formed through gas-rich mergers
has also been tested using simulations by \citet{bek05}, who showed
that the properties of K+A galaxies could be reproduced by merging two
gas-rich systems, although the details depend strongly upon the orbital
configuration. In addition, based on the kinematic study of K+A with
integral field unit (IFU) spectroscopy for 10 nearby K+A galaxies,
\citet{pra09} found that the majority of their K+A galaxies can be
classified as 'fast rotators', which is consistent with a product of
gas-rich mergers. However, whether our high-redshift K+A galaxies are subjected to the same mechanism that governs the low-redshift K+A formation is not well-understood \citep{yan09}. Although limited by small number statistics, our
results at $z\sim 1$ are consistent with the scenario where wet mergers could be
associated with the formation of K+A galaxies, and that mixed mergers
might also contribute to some fraction of K+As, as mixed mergers
together with wet mergers share similar environment distributions as
K+As. Whether mixed mergers are able to quench the star formation of
the gas-rich component and results in K+A phases will be an
interesting topic to investigate further in simulations.

\begin{deluxetable*}{lccc}
\tabletypesize{\scriptsize}
\tablewidth{0pt}
\tablecaption{Three Statistical Tests of Differences Between Samples.\label{tab:kstest}}
\tablehead{
    \colhead{Subsamples} &
    \colhead{Kolmogorov-Smirnov} &
    \colhead{Anderson-Darling} &
    \colhead{Mann-Whitney-Wilcoxon test}
}

\startdata
wet   vs. K+A	    &0.169   &0.354   &0.398   \\
dry   vs. K+A	    &0.085   &0.078   &0.039   \\
mixed vs. K+A	    &0.011   &0.079   &0.057   \\
wet+mixed vs. K+A   &0.172   &0.290   &0.243
\enddata

\tablecomments{For most statistical tests, the values given above are the p-value, which gives the probability of the null hypothesis that the two samples are drawn from the same population. Conventionally the threshold significance level of 0.05 is adopted to rule out the null hypothesis.}

\end{deluxetable*}

\subsection{The Role of Major Mergers in Forming Red-Sequence Galaxies}
It has been known that galaxy properties such as their colors,
morphologies, and star formation histories depend upon the environment
where they reside \citep[][also see Kauffmann et al. 2004 for their discussion on the influence of stellar mass in addition to the environment]{dre80,bla06,coo06,coo07,tas09}. For example,
the fraction of red, old, and S0/E types of galaxies increase in
higher density regions \citep{dre80}. One class of processes is the
so-called 'internal process', in which scenario galaxies evolve
passively without interactions with other galaxies or the surrounding
material such as IGM (Inter-Galactic Medium) or ICM (Intra-Cluster
Medium). In this model, the average age of galaxies is older in dense
regions simply because they have formed earlier than those in
underdense regions. Other types of mechanisms are driven by the
'external process', including galaxy mergers, ram-pressure stripping,
galaxy harassment, and strangulations as discussed in \S
\ref{sec:KA}. It has become clear that passive evolution alone can not
fully account for the change in the properties of galaxies across
various environments, in particular the morphological
transformations. Therefore the question is no longer whether the
galaxies are evolved through 'nature' (internal) or 'nurture'
(external) processes, but rather to what level do the external
processes contribute to the evolution of galaxies and which extrinsic
process is the dominant factor.

How does this paper relate to this subject? The derived fractional
merger rate we derive suggest that galaxy mergers play a
non-negligible role, in particular in dense environments. The higher
frequency of dry mergers occurring in dense environments
relative to underdense regions will lead to the following
consequences: the structure parameters and the stellar mass function
of red-sequence galaxies in dense environments should be different
from those in underdense regions. This is because dry mergers tend to
produce boxy, slowly rotating anisotropic systems \citep{kho05,naa06}, and
also increase the stellar mass per galaxy. The picture sketched above is in agreement with recent studies of the
stellar mass function showing that red galaxies in dense
environments are typically more massive than their counterparts in
underdense regions \citep{bun06,yang09,bol09}.

Living at the high-mass
extreme of the galaxy population in overdense environments are
Brightest Cluster Galaxies (BCGs).
In spite of the numerous studies of the properties of BCGs,
their origin and evolution remain an unresolved issue. While there
is evidence of on-going dry mergers found at the centers of
groups and clusters at low and
intermediate redshifts \citep{mul06,rin07,mci08,tra08,liu09}, analyses
based on stellar populations of BCGs indicate that BCGs have assembled
most of the stellar mass ($\sim$ 90\%) by at least $z \sim 1$, leaving
little room for hierarchical
assembling through mergers \citep{whi08,col09}. Whether the 38$\pm$10\%
mass accretion rate through dry mergers we derive in
high-density regions is compatible with the $\sim$ 10\% growth in the
typical stellar mass of BCGs between $z = 1$ and $z = 0$ depends on the
actual abundances of BCGs over this redshift range.
The number density of massive halos at $z \sim 1$ is much lower than that at $z \sim 0$ in $\Lambda$CDM models. If the number of BCGs correlates with the number of massive halos in a similar manner between low and high redshifts, it is expected that a significant fraction of the progenitors of present-day BCGs has not appeared in the form of BCGs at $z \sim 1$ yet. In other words, the simplest explanation to alleviate the aforementioned potential discrepancy is that $z
\sim 1$ BCGs only represent some portion of the present-day BCG population and
the rest formed via successive dry mergers over this period. We remark, however, that our dry merger samples are likely going to be sub-BCG systems as very massive clusters are rare given our survey volume. Surveys over larger areas are required to tackle this issue more robustly.

The picture uncovered by this work can be summarized as follows.  At $z
\sim 1$ wet mergers occur at a similar rate across different
environments ($\sim$ 0.06$\pm$0.01 per Gyr at $0.75 < z < 1.0$ and $\sim$ 0.12$\pm$0.01 per Gyr at $1.0 < z < 1.2$), while the frequency of dry and mixed
mergers increases with the local galaxy density. The latter
is primarily because of the increasing population of red galaxies with respect
to local densities.
As a consequence, while red-sequence galaxies in low-density
environments are mainly built up through wet mergers, the dry and
mixed mergers only become important at intermediate to
overdense environments.  Such events contribute to the assembly of massive red
galaxies and could signal the precursors of BCGs in clusters seen in the
local Universe. However, we caution that measurements of local density
may not always have a one-to-one correspondence with the physical
global environment (i.e., field, groups, and clusters). More accurate mapping of the relation
between merger rates and the growth of massive galaxies in galaxy
groups and clusters will have to await larger samples of interacting galaxies in those environments at low and high redshift.

\section{CONCLUSION}
Our results can be summarized as follows:

1. At $0.75 < z < 1.2$, the typical environment hosting mixed and dry mergers is denser than that of wet mergers, suggesting that the roles of wet, dry, and mixed mergers in the galaxy evolution vary with environment. The difference in the local density distribution of various types of mergers is in broad agreement with predictions based on the observed color-density relation. However, we noticed an excess of dry and mixed pairs compared to the above expectation toward overdense regions at a $\sim$ 2-$\sigma$ level, indicating that the red and blue galaxies are not uniformly populated and there exists clustering effect at very small scales in those overdense environments. \\

2. In the redshift range ($0.75 < z < 1.2$) we have probed, we find a strong dependence of observed galaxy companion rate (\Nc) on environment, which holds for all types of pairs (blue-blue, red-red, blue-red pairs). Although \Nc increases with over density, using N-body simulations, we found that the fraction of pairs that will actually merge decreases with the local density. This is predominant because of a more pronounced projection effect in dense environments compared to low-density regions.\\

3. After correcting the environment dependence of the fraction of merger in pairs ($C_{mg}$) and the merger time-scale ($T_{mg}$), we find a weak environment dependence of the fractional merger rate for wet mergers over the redshift range $0.75 < z < 1.2$. The probability of a blue galaxy to merge with another blue galaxy is about 0.06$\pm$0.01 per Gyr at $z \sim 0.85$ and 0.12$\pm$0.01 per Gyr at $z \sim 1.1$. On the other hand, the fractional dry and mixed merger rate increases rapidly with local density due to the increased population of red galaxies in denser environments. The fraction of dry merger per Gyr is estimated to be (16$\pm$4)\% in overdense regions.\\

4. We find that the environment distribution of wet mergers alone or wet+mixed mergers is indistinguishable from that of the K+A galaxies, suggesting a plausible link between K+A galaxies and wet and/or wet+mixed mergers.\\

5. While wet mergers transform galaxies from the blue cloud into the red sequence at a similar fractional rate across different environments, dry mergers are most effective in high-density regions. We estimate that dry mergers contribute to (38$\pm$10)\% mass accretion of massive red-sequence galaxies in overdense environments, such as BCGs in massive groups and clusters since $z \sim 1$. Based on our results, we therefore expect that the properties including structures and masses of red-sequence galaxies should be different between those in underdense regions and in overdenser regions since dry mergers are only important in dense environments. \\

\acknowledgments

We thank the anonymous referee for helpful comments to improve this paper. L. Lin would like to thank E. Barton, D. McIntosh, and C. Conselice for their
helpful discussions. The work is partially
supported by the National Science Council of Taiwan
under the grant NSC99-2112-M-001-003-MY3. T. Chiueh and H.-Y. Jian acknowledge the support of NSC grant NSC97-2628-M-002-008-MY3. DEEP2 has been supported by NSF grants AST-0808133 and AST-0806732. The DEEP2 Redshift Survey has been made possible through the dedicated efforts of the DEIMOS instrument team  at UC
Santa Cruz and support of the staff at Keck Observatory. This work is also based in part on observations obtained with MegaPrime/MegaCam, a joint project of CFHT and CEA/DAPNIA, at the Canada-France-Hawaii Telescope (CFHT) which is operated by the National Research Council (NRC) of Canada, the Institute National des Sciences de l'Univers of the Centre National de la Recherche Scientifique of France, and the University of Hawaii. Access to the CFHT was made possible by the Ministry of Education and the National Science Council of Taiwan as part of the Cosmology and Particle Astrophysics (CosPA) initiative. We close with thanks to the Hawaiian people for use of their sacred mountain.

\end{document}